\documentclass{article}
\usepackage[T1]{fontenc} 
\usepackage[utf8]{inputenc} 
\usepackage{ismir,amsmath,cite,url}
\usepackage{amssymb}
\usepackage{amsfonts}
\usepackage{graphicx}
\usepackage{color}

\usepackage{siunitx}

\usepackage{booktabs}  

\usepackage[american]{babel}
\usepackage[babel]{csquotes}

\newcommand{\topi}[1]{^{#1}}

\usepackage{enumitem}

\newcommand{\field}[1]{\mathbb{#1}}

\newcommand{\R}{\field{R}}

\title{A Study of Annotation and Alignment Accuracy for Performance Comparison in Complex Orchestral Music}
\multauthor
{Thassilo Gadermaier$^1$ \hspace{1cm} Gerhard Widmer$^{1,2}$ \hspace{1cm}} { 
 $^1$ Institute of Computational Perception, Johannes Kepler University Linz, Austria\\
 $^2$ Austrian Research Institute for Artificial Intelligence (OFAI), Vienna, Austria\\
{\tt\small thassilo.gadermaier@jku.at, gerhard.widmer@jku.at}
}
\def\authorname{Thassilo Gadermaier, Gerhard Widmer}

\begin{document}

\maketitle


\begin{abstract}
Quantitative analysis of commonalities and differences between recorded music performances is an increasingly common task in computational musicology.
A typical scenario involves manual annotation of different recordings of the same piece along the time dimension, for comparative analysis of, e.g., the musical tempo, or for mapping other performance-related information between performances.
This can be done by manually annotating one reference performance, and then automatically synchronizing other performances, using audio-to-audio alignment algorithms.
In this paper we address several questions related to those tasks.
First, we analyze different annotations of the same musical piece, quantifying timing deviations between the respective human annotators.
A statistical evaluation of the marker time stamps will provide (a) an estimate of the expected timing precision of human annotations and (b) a ground truth for subsequent automatic alignment experiments.
We then carry out a systematic evaluation of different audio features for audio-to-audio alignment, quantifying the degree of alignment accuracy that can be achieved, and relate this to the results from the annotation study.
\end{abstract}

\section{Introduction}\label{sec:introduction}

An increasingly common task in computational musicology --
specifically: music performance analysis -- consists in
annotating different performances (recordings) of classical
music pieces with structural information (e.g., beat positions)
that defines a temporal grid, in order then to carry out some
comparative performance analyses, which require time alignments
between the performances.
As manually annotating many recordings is a very time-consuming
and tedious task, an obvious shortcut would be to manually annotate
only one performance, and then use automatic audio-to-audio
matching algorithms to align additional recordings to it, and
thus also be able to automatically transfer structural annotations.

The work presented here is part of a larger project on the analysis of orchestral music performance. In this musicological context, it is crucial to understand the level of precision one can expect of the empirical data collected. The present study attempts to answer two specific questions: (1) what is the precision / consistency we can expect from human time annotations in such complex music? and (2) can automatic alignment be precise enough to be used for transferring annotations between recordings, instead of tediously annotating each recording manually? 
We will approach this by collecting manual annotations from expert musicians, on a small set of carefully selected pieces and recordings (Section \ref{sec:annotation}), analyzing these with statistical methods (Section \ref{sec:eval:annotations}) -- which will also supply us with a ground truth for the subsequent step --, then performing systematic experiments with different audio features and parameters for automatic audio-to-audio alignment (Section \ref{sec:alignment}), quantifying the degree of alignment precision that can be achieved, and relating this to the results from the previous annotation study (Section \ref{sec:eval:alignments}).

\section{Related Work}

\cite{weiss_2016_measure_annotation} presented a case study of opera recordings that were annotated by five annotators, at the bar level.

The authors used the mean values over the annotators as ground-truth values for the respective marker positions and the variance to identify sections possibly problematic to annotate, and offered a qualitative analysis of the musical material and sources for error and disagreement between annotators.

\cite{grachten_alignment_structure} deals with the alignment of recordings with possibly different structure.
Their contribution is relevant for our endeavor in so far as they evaluated different audio features and parameters ranges for an audio-to-audio alignment task on a data set of, among others, symphonies by Beethoven, which matches our data set very well.
\cite{kirchhoff_2011_evaluation_features_alignment} evaluated audio features for the audio-to-audio alignment task using several different data sets.

While many studies of alignment features do not use real human performances but artificial data, we only use ground-truth produced from human annotations (by averaging over multiple annotations per recording) of existing recordings for the evaluation of the alignment task.
Furthermore, the results of our analysis of manual annotations (Step 1) will inform our interpretation of the automatic alignment experiments in Step 2 (by relating the observed alignment errors to the variability within the human annotations), leading to some insights useful for quantitative musicological studies. 

\section{Annotation and Ground-truth}
\label{sec:annotation}

\subsection{Annotation vs.~Tapping}

\begin{figure*}
\centering
\includegraphics[width=1\textwidth]{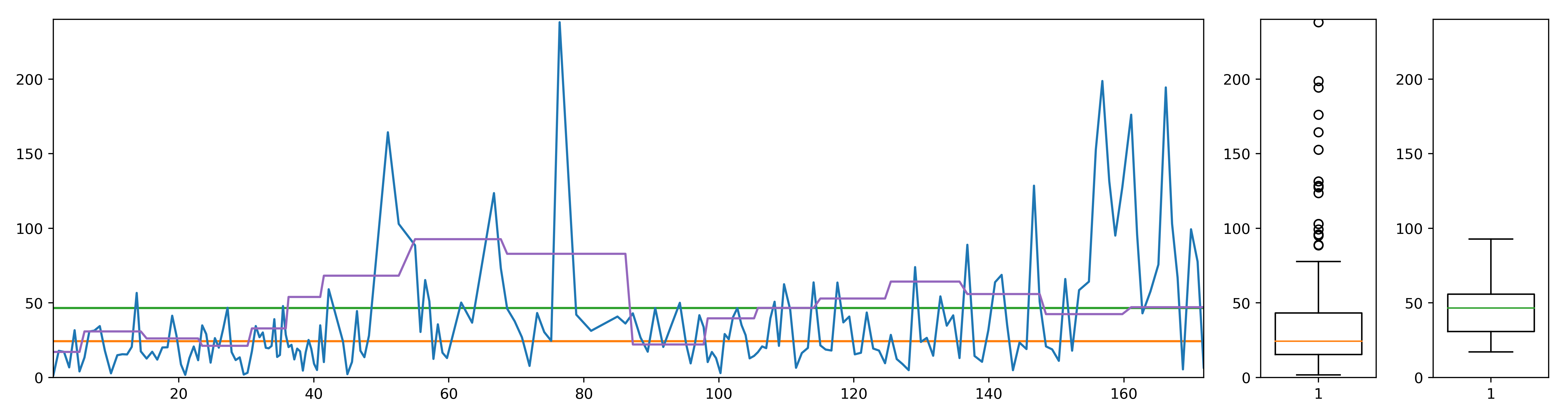}
\caption{
Standard deviations of annotations along a performance, Webern Op21-2, Boulez. Blue: computed from three markers per score event. Magenta: computed from pooled differences (details see text). Orange: median standard deviation (SD), green: median of SD from pooled differences (see boxplots right). Right: boxplots for SD (summary of blue curve) and for SD of pooled values (summary of magenta curve), central quartile is median.
} \label{fig:std_annotat}
\end{figure*}

Our primary goal is to map the musical time grid as defined by the score, to one or more performances given as audio recordings.
Due to expressiveness performance, these mapped time points may be very different between different recordings.
Following \cite{dixon_2005_match}, we will call the occurrence of one or more (simultaneous) score notes a \emph{score event}. 
In our case, we were interested in annotating regularly spaced score events, for instance, on the quarter note beats. 

Different methods can be employed for marking score events in a recording. 
One possibility is to tap along a recording on a keyboard (or other input device) and have the computer store the time-stamps.
We will refer to a sequence of time-stamps produced this way as a \emph{tapping} in the following.
Producing markers this way has been termed \enquote{reverse conducting} by the Mazurka project\footnote{\url{www.mazurka.org.uk/info/revcond/example/}}.

This is to be distinguished from what we will call an \emph{annotation} throughout this paper.
In that case, markers are first placed by tapping along, or even by visually inspecting the audio waveform, and then
iteratively corrected on (repeated) critical listening.
In general, we assume corrected annotations to have smaller deviations from the \enquote{true} time-stamps than uncorrected tappings, especially around changes of tempo.

\subsection{Pieces, Annotators, and Annotation Process}

The annotation work for this study was distributed over a pool of four annotators.
Three are graduates of musicology and one is a student of the violin.
The pieces considered are: Ludwig van Beethoven's Symphony No. 9, 1st movement; Anton Bruckner's Symphony No. 9, 3rd movement; and  Anton Webern's Symphony Op. 21, 2nd movement (see Table \ref{table:pieces} for details).

The first two are symphonic movements, played by full classical/romantic period orchestra.
The third is an atonal piece where the second movement is of a \enquote{theme and variations} form, and requires a much smaller ensemble (clarinets, horns, harp, string section).
While the first two pieces can be considered to be well known even to average listeners of classical music, the Webern piece was expected to be less familiar to the annotators. It is rhythmically quite complicated, with many changes in tempo and many sections ending in a fermata.
We expected it to be a suitable challenge for the annotators as well as the for the automatic alignment procedure.

The quarter beat level was chosen as (musically reasonable) temporal annotation granularity, in all three cases. The annotators were asked to mark
all score events (notes or pauses) at the quarter beat level, using the Sonic Visualiser software \cite{SonicVisualiser},
and then to correct markers such that they coincide with the score events when listening to the playback with a ``click" sound together with the recording of the piece.
They also had to annotate ``silent" beats (i.e. general pauses) or even single or multiple whole silent bars with the given granularity.
It is clear that this may create large deviations between annotators at such points, as the way to choose the marker positions is not always obvious or even meaningfully possible in these situations.

Each recording was annotated by three annotators, giving us a total of 21 complete manual annotations\footnote{Supplemental material to this publication is available online at 10.5281/zenodo.3260499}.

\begin{table}[]
\centering
\resizebox{\columnwidth}{!}{%
\begin{tabular}{@{}ccccc@{}}
\toprule
Composer        & Piece        & Part        & Section     &   \# Events   \\ \midrule
Beethoven       & Sym. 9       & 1st mov.    & complete    &    1093       \\
A. Bruckner     & Sym. 9       & 3rd mov.    & 150 - end   &    371        \\
A. Webern       & Sym. Op. 21  & 2nd mov.    & complete    &    198        \\ \bottomrule
\end{tabular}%
}
\caption{Annotated works/parts, and number of events. Granularity in all cases: quarter notes.}
\label{table:pieces}
\end{table}

\begin{table}[]
\small
\centering
{
\begin{tabular}{@{}cccccc@{}}
\toprule
Composer    & Conductor  & Orch. & Year & Dur.   &  Med. SD  \\ \midrule
Beethoven   & Karajan    & VPO   & '47  & 16:00  &  32       \\
            & Karajan    & BPO   & '62  & 15:28  &  32       \\
            & Karajan    & BPO   & '83  & 15:36  &  27       \\ \midrule
A. Bruckner & Karajan    & BPO   & '75  & 09:30  &  68       \\
            & Abbado     & VPO   & '96  & 10:40  &  52       \\ \midrule
A. Webern   & Boulez     & LSO   & '69  & 03:08  &  47       \\
            & Karajan    & BPO   & '74  & 03:28  &  63       \\ \bottomrule
\end{tabular}%
}
\caption{Annotated recordings. 
VPO = Vienna Philharmonic Orchestra, BPO = Berlin Philharmonic, LSO = London Symphony Orchestra.
Each recording was annotated by three annotators.
Med. SD is the median value of standard deviations of the annotations (in milliseconds, rounded to nearest integer), for details see Sec. \ref{sec:eval:annotations}.
} 
\label{table:recordings}
\end{table}

\section{Evaluation of Annotations}
\label{sec:eval:annotations}

For a statistical analysis of this rather small number of human annotations, we need to make some idealizing assumptions.
We assume that there is one clear point in time that can be attributed to each respective score event, i.e. there are \enquote{true} time-stamps $\tau_n$, $n = 1, 2, \ldots$ for the score events we sought to annotate.
If each score event is annotated multiple times, the annotated markers $\theta_n$ will show random variation around their true time-stamps, with a certain variance $\sigma^2_n$.
It seems reasonable to assume the respective markers to be realizations of random variables $\varTheta_n$, each following a normal distribution, i.e. $\varTheta_n \sim \mathcal{N}(\tau_n, \sigma_n^2)$.

\begin{figure}
\centering
\includegraphics[width=\columnwidth]{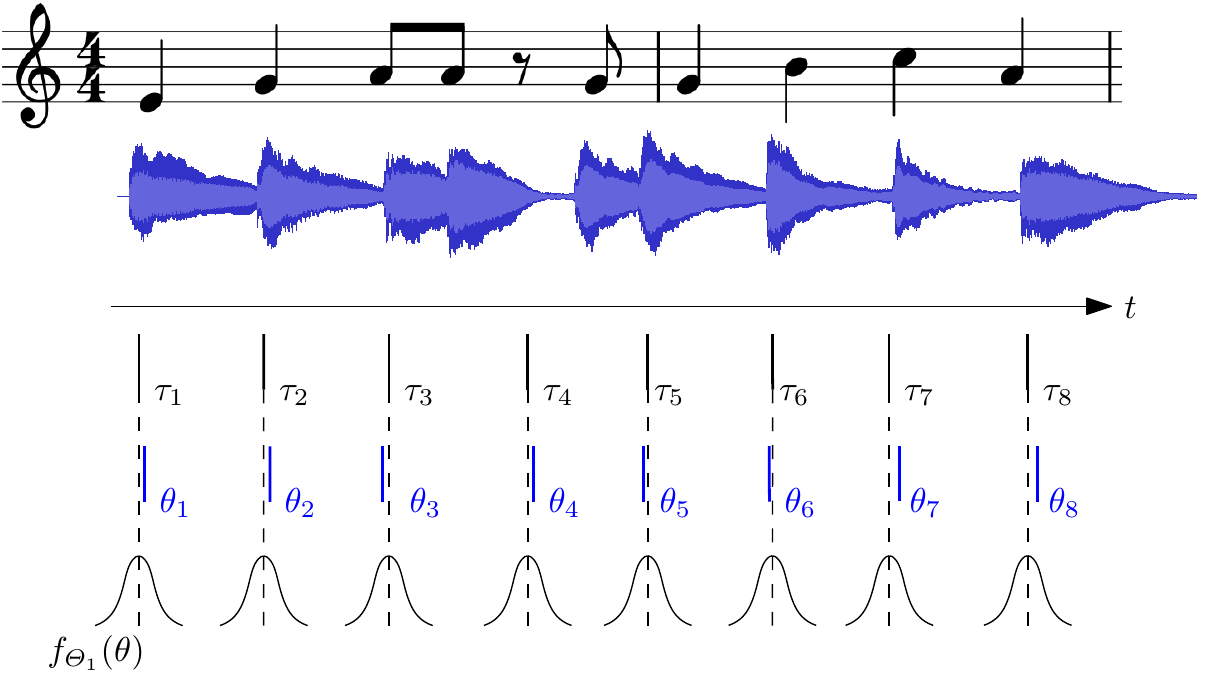}
\caption{
Modeling annotations as random variables.
Musical score and waveform of a performance.
Hypothetical true time-stamps $\tau_n$.
Annotation markers $\theta_n$.
Bottom row: pdfs of random variables $\varTheta_n$, each of mean $\tau_n$.
} \label{fig:theta_Theta}
\end{figure}

Thus, for each event to be annotated we would expect (a large number of) annotations to exhibit a normal distribution around some mean $\tau_n$.
This is schematically illustrated in Figure \ref{fig:theta_Theta}.

However, for estimating the parameters of these distributions, rather large numbers of annotations would be required.

\cite{dannenberg_2009_single_tap_error_distribution} has shown that with some additional assumptions, the distribution can be estimated from as little as two sequences of markers.

We follow \cite{dannenberg_2009_single_tap_error_distribution} in the derivations below.
If the variance $\sigma^2$ of the time stamps is assumed to be constant over time (across the whole piece or part to be annotated), subtracting two sequences $\theta_n^1$, $\theta_n^2$ of markers for the same score events, i.e.
\begin{equation} \label{eq:delta_diff}
\Delta_n = \varTheta_n\topi{1} - \varTheta_n\topi{2},
\end{equation}
yields the variable $\Delta_n \sim \mathcal{N}(0, 2 \sigma_\varTheta^2)$.
Note that if the mean of $\Delta_n$ is not zero, we can force it to be by suitably offsetting either $\varTheta_n\topi{1}$ or $\varTheta_n\topi{2}$ by $\bar{\Delta}_n$ -- since we assume both sequences to mark the same events with mean zero, a total mean deviation can be viewed as a systematic offset by either annotator.
One could then use the differences $\delta_n = \theta_n^1 - \theta_n^2$ to estimate the variance $\sigma_\varTheta^2$ around the true time-stamps:
\begin{equation} \label{eq:estimate_var}
    \hat{\sigma}_\varTheta^2 = \frac{1}{2N}\sum_{n = 1}^N (\theta_n\topi{1} - \theta_n\topi{2})^2.
\end{equation}
In \cite{dannenberg_2009_single_tap_error_distribution}, two example analyses of tap sequences were presented that support these assumptions.

\begin{figure}
\centering
\includegraphics[width=1\columnwidth]{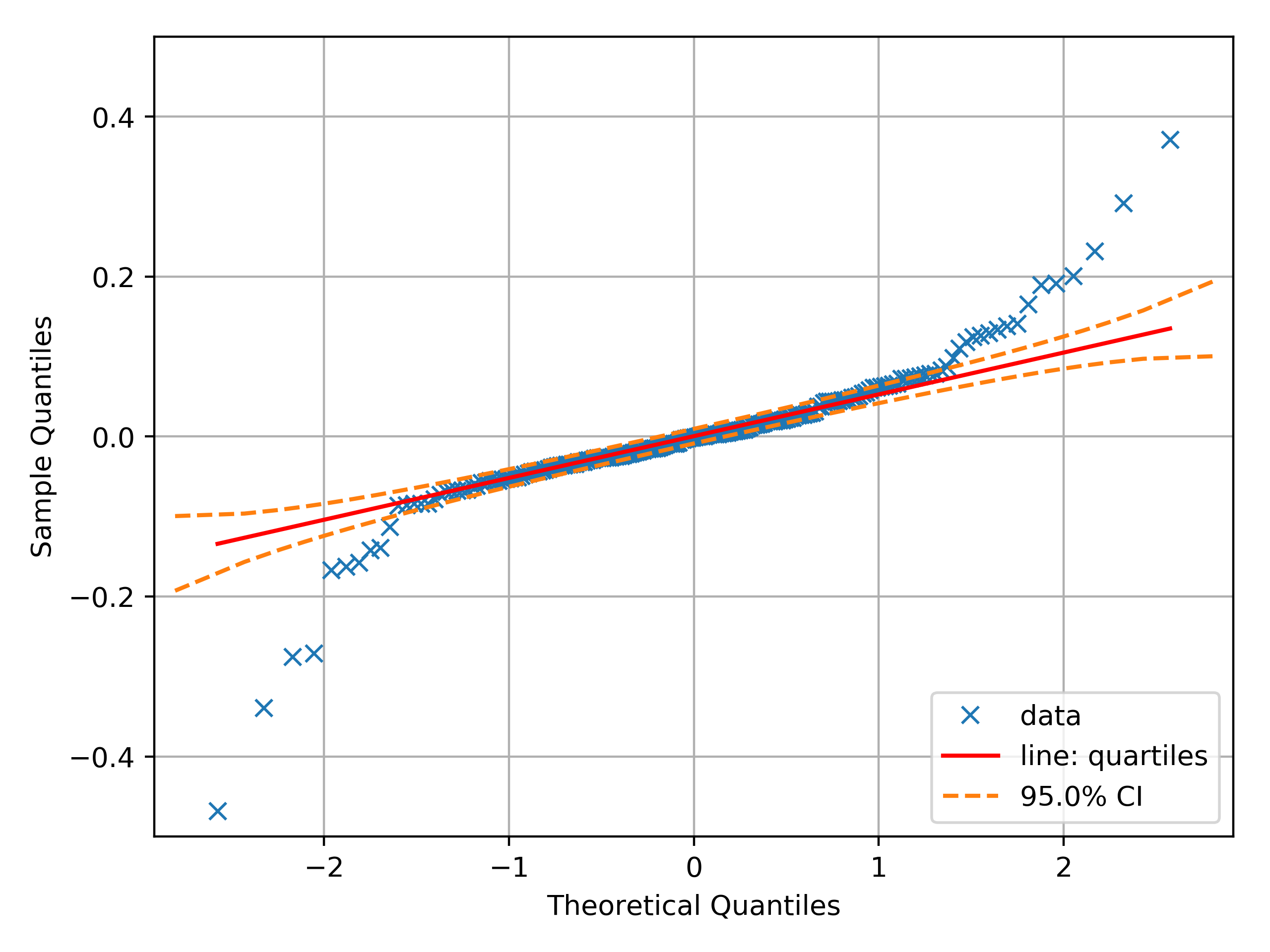}
\caption{
Webern Op21-2, Boulez. Quantile-quantile plot of the differences of a pair of annotation sequences for the whole piece. Solid red line fitted to first and third data quartile, dashed lines show $\pm$95\% confidence around this line. Non-normal data deviate strongly from area enclosed by dashed lines. 
} \label{fig:qqplot_webern}
\end{figure}

We analyzed our annotation data according to these ideas.
First, for each annotated recording, we calculated the time-stamp differences between each pair of annotations, according to Eq. \eqref{eq:delta_diff}, and tested the resulting distributions for normality, using the Shapiro-Wilk test.
However, for all annotations created, none of the distributions is normal according to these tests. 
On visual inspection of the distributions of differences of annotation sequences $\delta_n$ using quantile-quantile plots (see Fig. \ref{fig:qqplot_webern}), the tails of the distributions turned out to be typically significantly heavier than for a normal distribution.

We suspect that this discrepancy to the results given in  \cite{dannenberg_2009_single_tap_error_distribution} is most likely due to the higher complexity of our musical material, with large orchestras playing highly polyphonic and rhythmically complex music in varying tempi.
It seems intuitively clear that for some sections, the deviations among annotated markers will be much smaller than in complex parts.
Additionally, as we asked also silent beats to be annotated, even during whole silent bars, we should expect substantial deviations for at least a few such events in every recording.
We therefore conclude that at least the assumption of identical variance across a whole piece should be dropped (for more complex material) when more detailed information about local uncertainties of the annotation is desired.

However, it is interesting to note that locally, when the differences for only a few consecutive (around 20-30) annotated time-stamps are pooled, they conform to a normal distribution quite well.
This means that the assumption of about equal variance for the annotation of score events tends to hold for short blocks of time, but rather not globally (for a whole piece), at least for the musical material considered here.

As estimating the standard deviation (as a measure of uncertainty) of each time-stamp's markers is not reliable given only few annotations, we used an alternative based on the above observation.
For blocks of 24 consecutive score events (with a hop size of 12), the differences of a pair of annotation sequences were \textit{pooled} and used to estimate the standard deviation for each respective block.
The resulting, block-wise constant curve of standard deviations is shown in Fig. \ref{fig:std_annotat} (magenta), along with the simple standard deviation per score event, calculated from three markers (blue), for a specific recording and pair of annotations.
The median of these per-block estimated standard deviations is used as a global estimate of the precision of the annotations for the respective performance, and is given for the respective performance as the right-most column in Table \ref{table:recordings}.
As can be seen, the values differ substantially across the pieces as well as within the pieces, for different performances.
The right-most boxplot in Fig. \ref{fig:std_annotat} shows a summarization of the per-block estimated standard deviations.
Interestingly, for the 1st movement of Beethoven's Symphony 9 (with its relatively constant tempo), the estimated standard deviation is close to the value presented in \cite{dannenberg_2009_single_tap_error_distribution}, but it is considerably larger for the other pieces that exhibit more strongly varying tempo.

\section{Automatic Alignment}
\label{sec:alignment}

As mentioned above, annotating a large number of performances of the same piece is a time-consuming process.
A more efficient alternative would be to automatically transfer annotations from one recording to a number of unseen recordings, via audio-to-audio alignment.

\subsection{Alignment Procedure and Ground-truth}

\begin{figure}
\centering
\includegraphics[width=0.5\columnwidth]{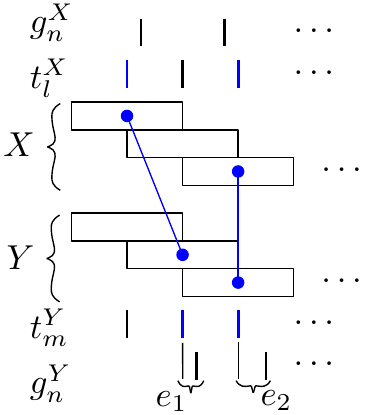}
\caption{
Matching feature vectors through DTW, and calculating errors between associated time-stamps $t_m^Y$ and ground-truth time-stamps $g_n^Y$, for direction $X \rightarrow Y$.
This yields the error sequence $e_n^{X \rightarrow Y}$.
} \label{fig:alig_ts_gt}
\end{figure}

The method of choice for (off-line) audio-to-audio alignment is \textit{Dynamic Time-warping (DTW)} \cite{muller_2019_cross_modal}.
Aligning two recordings via DTW involves extracting sequences $X \in \R^{L \times D}$ and $Y \in \R^{M \times D}$ of feature vectors, respectively.
Using a distance function $d(x_l, y_m)$, the DTW algorithm finds a path of minimum cost, i.e. a mapping between elements $x_l$, $y_m$ of the sequences $X$, $Y$.
An alignment is thus a mapping between pairs of feature vectors (from different recordings), each vector representing a block of consecutive audio samples.
As each audio sample has an associated time-stamp (an integer multiple of the inverse of the sample rate), each feature vector, say $x_l$, can be associated with a time-stamp $t_l^X$ as well, (here) representing the center of the block of audio samples.
The matching of sequence elements is schematically illustrated in Fig.~\ref{fig:alig_ts_gt}, for the \enquote{direction} $X \rightarrow Y$ (note that direction here refers to the evaluation, as will be illustrated next).
For each block of $X$, the matching block of $Y$ is found, and its associated time-stamp $t_m^Y$ is subtracted from the ground-truth time-stamp $g_n^Y$.
This produces the pairwise error sequence $e_n^{X\rightarrow Y}$.
As we have ground-truth annotations for both recordings of a pair available, we can also calculate an error sequence for the \enquote{reverse} direction $Y \rightarrow X$.

The sequences of ground-truth time-stamps were produced from the multiple annotations discussed above (Section \ref{sec:annotation}), by taking for each annotated score event the sample mean across the three annotations per recording.
For computing the alignments, an implementation of FastDTW \cite{salvador_2007_fastdtw} in python was used.

\subsection{Choice of Audio Features}

The actual alignment process is preceded by extracting features from the recordings to be aligned.
Different features have been proposed and evaluated for this task in the literature.
We decided to choose only features that have proven to yield highly accurate alignments and thus small alignment errors.

\cite{grachten_alignment_structure} evaluated several different audio features separately on data sets of different music genres, among them symphonies by Beethoven.
They achieved the best results overall by using 50 MFCC (in contrast to 13 or even 5 as used in \cite{kirchhoff_2011_evaluation_features_alignment}), for two different block lengths.
As the results on these corpora, which are similar to ours, were dominated by MFCC, we decided to use these with similar configurations for our experiments.
Additionally, we included a variant of MFCC (in the following addressed as \enquote{MFCC mod}) following an idea described in \cite{muller_2009_chroma_features_}, where 120 MFCC are extracted, then the first $n_{skip}$ are discarded and only the remaining ones used.
However, in contrast to their proposal we skip the subsequent extraction of chroma information and use the MFCC directly.

The second family of features that has proven successful for alignment tasks are chroma features, which were tested as an alternative.
For extracting the feature values, the implementations from LibROSA \cite{librosa_2015} were used.
Besides \enquote{classical} chroma features, the variants chroma\_cqt (employing a constant-Q transform) and chroma\_cens were used.
We decided not to include more specialized features that include local onset information, like LNSO / NC \cite{arzt_2012_adaptive_distance}, or DLNCO (in combination with chroma), as they would seem to give no advantage on our corpus as suggested from the results in \cite{grachten_alignment_structure} and \cite{ewert_2009_hires_sync_chroma}.

\subsection{Systematic Experiments Performed}

In order to find the best setup for audio-to-audio alignment for complex orchestral music recordings, we carried out a large number of alignment experiments, by systematically varying the following parameters:
\begin{itemize} [nolistsep]
    \item FFT sizes: 1024 to 8192 (chroma), up to 16384 (MFCC)
    \item Hop sizes MFCC: half of FFT size, for 16384 fixed to 4096
    \item Hop sizes Chroma: 512 and 1024, for each FFT size; additionally 2048 for chroma\_cens and chroma\_cqt
    \item Number of MFCC: 13, 20, 30, 40, 50, 80, 100
    \item MFCC mod: 120 coefficients, first 10, 20, $\ldots$, 80 discarded
    \item Distance measures: Euclidean ($l_2$), city block ($l_1$) and cosine distance.
\end{itemize}
Note that the audio signals were not down-sampled in any of the cases, but used with their full sample rate of 44.1 kHz.

All in all, a total number of 312 different alignments were computed and evaluated for each performance pair.
Each alignment of each pair of performances was evaluated in both directions.
As it is impossible to display all results in this paper, we will only report a subset of best results in Section \ref{sec:eval:alignments}.

\section{Evaluation of Alignments}
\label{sec:eval:alignments}

\subsection{Alignment Accuracy}

For quantifying the alignment accuracy, we calculated pairwise errors $e_n$ between the ground-truth time-stamps $g_n$ for the respective recording and the matching alignment time-stamps $t_l$ (see Fig. \ref{fig:alig_ts_gt}).
Per pair of recordings, two error sequences are obtained, one for each evaluation direction, i.e. $e_n^{X \rightarrow Y}$ and  $e_n^{Y \rightarrow X}$.
As a general global measure of the accuracy of a full alignment, the mean absolute error is used, where the maximum absolute error can be seen as a measure for lack of robustness.

\begin{figure}[htbp]
\centering
\includegraphics[width=\linewidth]{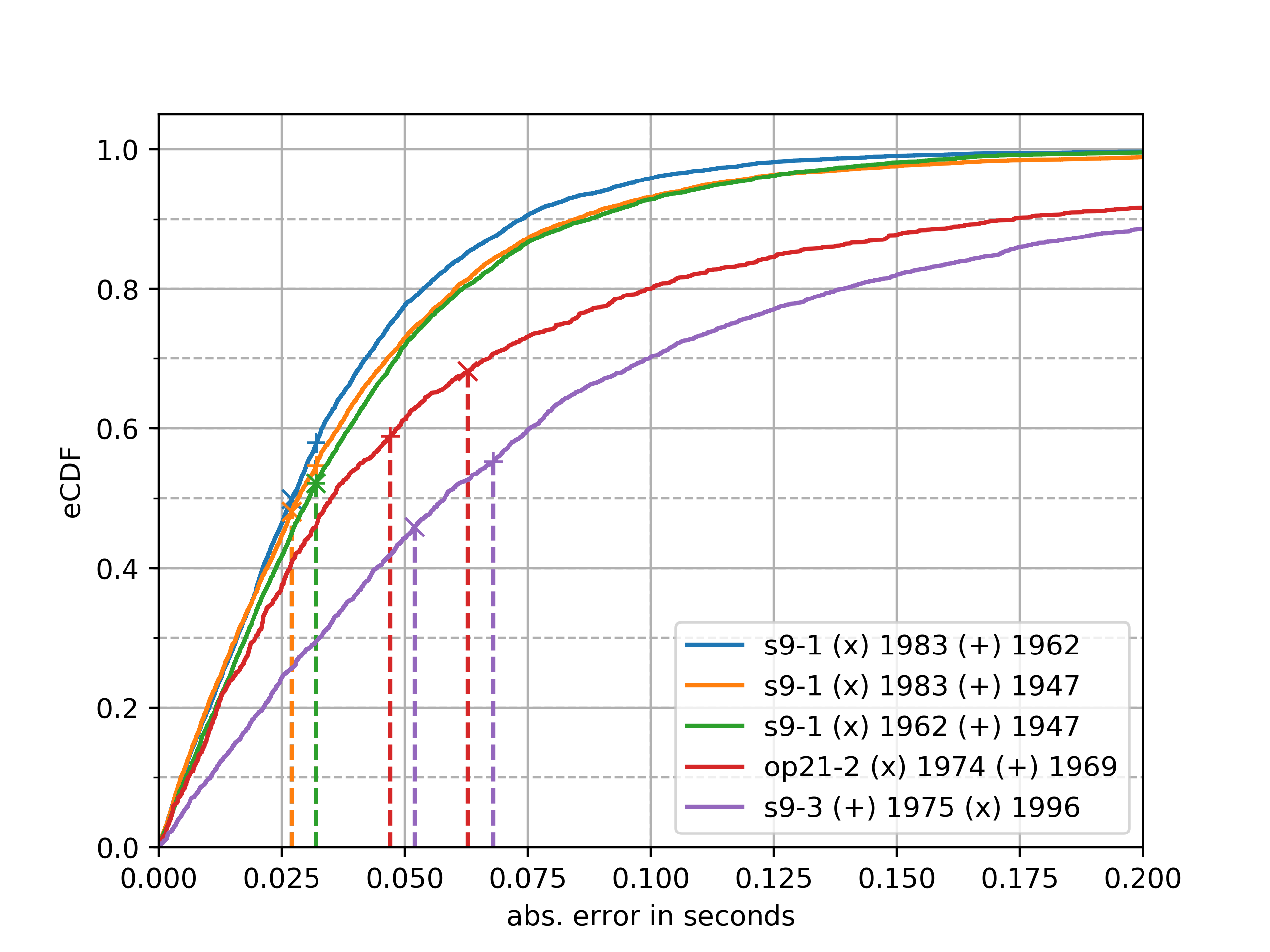}
\caption{Cumulative distribution of absolute pairwise errors. Each curve represents pooled errors of 10 best alignments (mean absolute error) for both evaluation directions, per pair of recordings. s9-1: Beethoven, S.9, 1st Mov., s9-3: Bruckner, S.9, 3rd Mov., op21-2: Webern, Op21, 2nd Mov. (+) and (x) markers for median standard deviation of annotation, cf. Table \ref{table:recordings}.}
\label{fig:alig_eCDF_summarized}
\end{figure}

For reporting of the best results, we first ranked all alignments whose absolute maximum errors are below 5 seconds by their mean absolute errors.
As large maximum error is taken as lack of robustness, the worst performing settings were thus discarded.
For each pair of recordings, from the remaining error sequences (from originally 312 alignments per pair, each with 2 directions of evaluation), the 10 best results, in terms of mean absolute error, were then kept for further analysis.
The error values for both directions of each specific alignment were then pooled, i.e. the error values were collected and analyzed jointly.
A one-way ANOVA (null hypothesis: no difference in the means) was conducted for the 10 best alignments per pair of recordings, where for all cases the null hypothesis could not be rejected (recording pair with smallest p-value: $F = 0.6$, $p = 0.8$).
Thus, as the different settings of the 10 best alignments do not result in significant differences in terms of mean error performance, the error sequences for those 10 best alignments were collected, to estimate a distribution of the absolute errors.
Fig. \ref{fig:alig_eCDF_summarized} shows the empirical cumulative distribution function of the pairwise absolute errors for all 5 alignment (performance) pairs, where each curve is obtained from the 2 error sequences (both evaluation directions) of each of the 10 best alignments for the respective performance pair.

In the following, the settings and results, in terms of mean absolute error and maximum absolute error, for the 10 best alignments are presented.
For the Beethoven piece, we restricted the reporting to one pair of recordings (BPO 1962 vs. VPO 1947) due to limited space (Table \ref{table:results_beethoven}).
As can be seen from Fig. \ref{fig:alig_eCDF_summarized}, the other two pairs do not differ substantially in terms of error performance, and the settings for obtaining these results are almost identical to the ones presented in the table, with an even stronger favor of the MFCC mod feature.
Tables \ref{table:results_webern} and \ref{table:results_bruckner} show the results for the Webern and Bruckner pair of recordings, respectively.

\begin{table}[]
\centering
\resizebox{\columnwidth}{!}{%
\begin{tabular}{@{}ccccccc@{}}
\toprule
Feature    & \#MFCC    & \#skip   & fft size    & dist.     & mean err. & max. err   \\ \midrule
MFCC mod   &  120      & 20       & 2048     &    $l_1$     &    38       &  220         \\
MFCC mod   &  120      & 30       & 4096     &    $l_2$     &    38       &  411        \\
MFCC mod   &  120      & 40       & 2048     &    $l_2$     &    39       &  239          \\
MFCC       &  100      & -        & 8192     &    $l_1$     &    40       &  243          \\
MFCC       &   80      & -        & 2048     &    $l_1$     &    40       &  253          \\
MFCC mod   &  120      & 10       & 4096     &    $l_1$     &    41       &  318          \\
MFCC       &  100      & -        & 16384    &    $l_1$     &    41       &  318          \\
MFCC mod   &  120      & 10       & 2048     &    $l_2$     &    41       &  370          \\
MFCC mod   &  120      & 20       & 16384    &    $l_1$     &    42       &  285          \\
MFCC mod   &  120      & 20       & 16384    &    cos       &    43       &  709          \\ \bottomrule
\end{tabular}%
}
\caption{Settings and results for top 10 alignments, Beethoven S9-1, BPO 1962 vs. VPO 1947. The other two cases show almost identical results (omitted for lack of space), with stronger favor of MFCC mod. Errors in ms, rounded to nearest integer.}
\label{table:results_beethoven}
\end{table}
\begin{table}[]
\centering
\resizebox{\columnwidth}{!}{%
\begin{tabular}{@{}ccccccc@{}}
\toprule
Feature    & \#MFCC    & \#skip   &fft size    & dist.     & mean err. & max. err   \\ \midrule
MFCC mod   &  120      & 30       & 2048    &    cos       &    116       &  4133         \\
MFCC mod   &  120      & 20       & 4096    &    cos       &    123       &  3901        \\
MFCC mod   &  120      & 50       & 2048    &    $l_1$     &    127       &  3341          \\
MFCC mod   &  120      & 40       & 8192    &    cos       &    137       &  3597          \\
MFCC mod   &  120      & 40       & 2048    &    cos       &    138       &  4180          \\
MFCC mod   &  120      & 60       & 4096    &    $l_2$     &    139       &  2639          \\
MFCC mod   &  120      & 10       & 16384   &    cos       &    144       &  4319          \\
MFCC mod   &  120      & 20       & 2048    &    cos       &    145       &  4110          \\
MFCC mod   &  120      & 20       & 16384   &    cos       &    150       &  4226          \\
MFCC mod   &  120      & 60       & 16384   &    cos       &    150       &  4040          \\ \bottomrule
\end{tabular}%
}
\caption{Settings and results for top 10 alignments, Bruckner S9-3. Errors in ms, rounded to nearest integer.}
\label{table:results_bruckner}
\end{table}
\begin{table}[]
\centering
\resizebox{\columnwidth}{!}{%
\begin{tabular}{@{}ccccccc@{}}
\toprule
Feature    & \#MFCC    & \#skip   & fft size     & dist.    & mean err.  & max. err   \\ \midrule
MFCC       &  80       & -         & 2048    &    $l_1$    &    62       &  1049     \\
MFCC       &  40       & -         & 2048    &    $l_1$    &    65       &  1026     \\
MFCC       &  100      & -         & 4096    &    $l_1$    &    67       &  980      \\
MFCC       &  100      & -         & 4096    &    $l_2$    &    69       &  980      \\
MFCC mod   &  120      & 10        & 4096    &    $l_2$    &    69       &  980      \\
MFCC mod   &  120      & 20        & 4096    &    $l_2$    &    73       &  980      \\
MFCC       &  100      & -         & 4096    &    cos      &    76       &  980      \\
MFCC       &  80       & -         & 2048    &    cos      &    77       &  980      \\
MFCC       &  100      & -         & 8192    &    $l_1$    &    78       &  1026     \\
MFCC mod   &  120      & 20        & 2048    &    cos      &    82       &  956      \\ \bottomrule
\end{tabular}%
}
\caption{Settings and results for top 10 alignments, Webern Op.21-2. Errors in ms, rounded to nearest integer.}
\label{table:results_webern}
\end{table}

As can be seen from the tables, best results are achieved with either MFCC or the modified MFCC.
There does not seem to be a very clear pattern of which parameter setting gives best results, even within one pair of recordings.
A slight advantage of medium to large FFT sizes is observed, as is a larger number of MFCC ($\geq$ 80, a number much larger than what is suggested in the literature for timbre related tasks). %
For the modified MFCC, skipping the first 20 to 40 out of the 120 coefficients seems a good suggestion.
Interestingly, there seems to be no clear relation to the FFT size. %

\subsection{Relation to Human Alignment Precision}

We would like to relate the accuracy achieved by automatic alignment methods to the precision with which human annotators mark score events in such recordings.
This will enable us to judge the errors in the alignment methods in such a way that we cannot only say which is best, but which are probably sufficiently good for musicological studies (in relation to how precise human annotations tend to be).

By comparing the global measures of variation of the annotations (Table \ref{table:recordings}) with the mean errors obtained from the alignment study, the following can be stated.
We would like the errors introduced by the alignments to be in the range of the variation introduced by human annotators.
If, for example, the above estimated standard deviations are used for describing an interval (e.g. $\pm$ 1 SD) around the ground-truth annotations, then markers placed by the DTW alignment within such an interval can be taken to be as accurate as an average human annotation.
However, as Tables \ref{table:results_beethoven} to \ref{table:results_webern} reveal, on average, the absolute errors are at least slightly (or even much in case of the Bruckner performances) larger than the estimated standard deviations, but still in a reasonable range, even for larger proportions of the score events (see Figure \ref{fig:alig_eCDF_summarized}).

\section{Discussion and Conclusions}

Given our results, we expect the presented feature settings to be quite suitable as a first step for developing further musicological questions related to comparing multiple performances of one piece.
With careful annotation of one recording, transferring the score event markers to other recordings of the same piece should yield not much worse accuracy than what is to be expected from human annotations.
Detailed analyses of e.g. tempo may still need a moderate amount of manual correction, however.

An interesting application we consider is the exploration of a larger corpus of unseen recordings.
Being able to establish, within a reasonable uncertainty, a common musical grid for a number of performances allows for search of (a first impression of) commonalities and differences across performances, for parameters such as tempo, or features extracted directly from the recording, such as loudness, mapped to the musical grid.
This will e.g. allow the pre-selection of certain performances for more careful human annotation and further more detailed analyses.
Recently, performance related data have been presented for a larger corpus in \cite{kosta_2018_mazurkabl}.

We hope to have presented some new insights with the data on annotation precision, and the applied methods for their quantification.
Further work could make use of estimates of typical uncertainty of annotations to estimate, or give bounds for, the uncertainty of data derived from these.
One way would be to use simple error propagation to quantify uncertainty of tempo representations, and automatically find (sections of) performances of significantly different tempo within a large corpus of recordings.


\section{Acknowledgments}
This work was supported by the Austrian Science Fund (FWF) under project number P29840, and by the European
Research Council via ERC Grant Agreement 670035, project CON ESPRESSIONE.
We would like to thank the annotators for their work, as well as the anonymous reviewers for their valuable feedback.
Special thanks go to Martin Gasser for fruitful discussions of an earlier draft of this work.

\bibliography{ismir_2019_ARXIV_literature.bib}

\end{document}